# Ejecta from the DART-produced active asteroid Dimorphos

Jian-Yang Li[1], Masatoshi Hirabayashi[2], Tony L. Farnham[3], Jessica M. Sunshine[3], Matthew M. Knight[4], Gonzalo Tancredi[5], Fernando Moreno[6], Brian Murphy[7], Cyrielle Opitom[7], Steve Chesley[8], Daniel J. Scheeres[9], Cristina A. Thomas[10], Eugene G. Fahnestock[8], Andrew F. Cheng[11], Linda Dressel[12], Carolyn M. Ernst[11], Fabio Ferrari[13], Alan Fitzsimmons[14], Simone Ieva[15], Stavro L. Ivanovski[16], Teddy Kareta[17], Ludmilla Kolokolova[3], Tim Lister[18], Sabina D. Raducan[19], Andrew S. Rivkin[11], Alessandro Rossi[20], Stefania Soldini[21], Angela M. Stickle[11], Alison Vick[12], Jean-Baptiste Vincent[23], Harold A. Weaver[11], Stefano Bagnulo[24], Michele T. Bannister[25], Saverio Cambioni[26], Adriano Campo Bagatin[27, 28], Nancy L. Chabot[11], Gabriele Cremonese[29], R. Terik Daly[11], Elisabetta Dotto[15], David A. Glenar[30], Mikael Granvik[31, 32], Pedro H. Hasselmann[15], Isabel Herreros[33], Seth Jacobson[34], Martin Jutzi[19], Tomas Kohout[35, 36], Fiorangela La Forgia[37], Monica Lazzarin[37], Zhong-Yi Lin[38], Ramin Lolachi[39, 40], Alice Lucchetti[29], Rahil Makadia[41], Elena Mazzotta Epifani[15], Patrick Michel[42], Alessandra Migliorini[43], Nicholas A. Moskovitz[17], Jens Ormö[33], Maurizio Pajola[29], Paul Sánchez[44], Stephen R. Schwartz[1], Colin Snodgrass[7], Jordan Steckloff[1], Timothy J. Stubbs[40], Josep M. Trigo-Rodríguez[45]

**Author Affiliations**

1. Planetary Science Institute, Tucson, AZ, USA
2. Auburn University, Auburn, AL, USA
3. Department of Astronomy, University of Maryland, College Park, MD, USA
4. United States Naval Academy, Annapolis, MD, USA
5. Departamento de Astronomía, Facultad de Ciencias, Udelar, Uruguay
6. Instituto de Astrofísica de Andalucía, CSIC, Glorieta de la Astronomía s/n, E-18008 Granada, Spain
7. University of Edinburgh, Royal Observatory, Edinburgh, UK
8. Jet Propulsion Laboratory, California Institute of Technology, Pasadena, CA, USA
9. University of Colorado, Boulder, CO, USA
10. Northern Arizona University, Flagstaff, AZ, USA
11. Johns Hopkins University Applied Physics Laboratory, Laurel, MD, USA
12. Space Telescope Science Institute, Baltimore, MD, USA
13. Department of Aerospace Science and Technology, Politecnico di Milano, Milano, Italy
14. School of Mathematics and Physics, Queen's University Belfast, Belfast, UK
15. INAF-Osservatorio Astronomico di Roma, Monte Porzio Catone, Roma, Italy
16. INAF - Osservatorio Astronomico di Trieste, Via G.B. Tiepolo, 11, Trieste, Italy
17. Lowell Observatory, Flagstaff, AZ, USA; Lunar and Planetary Laboratory, University of Arizona, Tucson, AZ, USA
18. Las Cumbres Observatory, Goleta, CA, USA
19. Space Research and Planetary Sciences, Physikalisches Institut, University of Bern, Bern, Switzerland

20. IFAC-CNR, Via Madonna del Piano 10, 50142, Sesto Fiorentino, Italy
21. Department of Mechanical, Materials and Aerospace Engineering, University of Liverpool, Liverpool, UK
22. Space Telescope Science Institute, Baltimore, MD, USA
23. DLR Institute of Planetary Research, Rutherfordstrasse 2, 12489 Berlin, Germany
24. Armagh Observatory and Planetarium, College Hill, Armagh BT61 9DG, United Kingdom
25. School of Physical and Chemical Sciences — Te Kura Matū, University of Canterbury, Private Bag 4800, Christchurch 8140, New Zealand
26. Department of Earth, Atmospheric and Planetary Sciences, Massachusetts Institute of Technology, Cambridge, MA 02139, USA
27. Instituto de Física Aplicada a las Ciencias y las Tecnologías. Universidad de Alicante, Spain
28. Departamento de Física, Ingeniería de Sistemas y Teoría de la Señal. Universidad de Alicante, Spain
29. INAF - Osservatorio Astronomico di Padova, vicolo Osservatorio 5, 35122 Padova, Italy
30. Center for Space Science and Technology, University of Maryland, Baltimore County, Baltimore, MD 21218
31. Department of Physics, University of Helsinki, Helsinki, Finland
32. Asteroid Engineering Laboratory, Luleå University of Technology, Kiruna, Sweden
33. Centro de Astrobiología (CAB), CSIC-INTA, Carretera de Ajalvir km4, 28850 Torrejón de Ardoz, Madrid, Spain
34. Department of Earth and Environmental Sciences, Michigan State University, East Lansing, MI, USA
35. Institute of Geology of the Czech Academy of Sciences, Prague, Czech Republic
36. Department of Geosciences and Geography, University of Helsinki, Finland
37. Dipartimento di Fisica e, Astronomia-Padova University, Vicolo dell'Osservatorio 3, 35122, Padova, Italy
38. Institute of Astronomy, National Central University, No. 300, Zhongda Rd., Zhongli Dist., Taoyuan City 32001, Taiwan
39. Center for Space Science and Technology, University of Maryland, Baltimore County, Baltimore, MD USA
40. NASA Goddard Space Flight Center, Greenbelt, MD, USA
41. Department of Aerospace Engineering, University of Illinois at Urbana-Champaign, Urbana, IL, USA
42. Université Côte d'Azur, Observatoire de la Côte d'Azur, CNRS, Laboratoire Lagrange, Nice, France
43. INAF - Institute of Space Astrophysics and Planetology, via Fosso del Cavaliere 100, 00133, Roma, Italy
44. Aerospace Engineering Sciences, Colorado Center for Astrodynamics Research, University of Colorado, Boulder, CO 80303, USA
45. Institute of Space Sciences (CSIC-IEEC), Campus UAB Bellaterra, 08193 Cerdanyola del Vallés (Barcelona), Catalonia, Spain

**Some active asteroids have been proposed to be the result of impact events[1]. Because active asteroids are generally discovered serendipitously only after their tail formation, the process of the impact ejecta evolving into a tail has never been directly observed. NASA's Double Asteroid Redirection Test (DART) mission[2], apart from having successfully changed the orbital period of Dimorphos[3], demonstrated the activation process of an asteroid from an impact under precisely known impact conditions. Here we report the observations of the DART impact ejecta with the Hubble Space Telescope (HST) from impact time T+15 minutes to T+18.5 days at spatial resolutions of ~2.1 km per pixel. Our observations reveal a complex evolution of ejecta, which is first dominated by the gravitational interaction between the Didymos binary system and the ejected dust and later by solar radiation pressure. The lowest-speed ejecta dispersed via a sustained tail that displayed a consistent morphology with previously observed asteroid tails thought to be produced by impact[4,5]. The ejecta evolution following DART's controlled impact experiment thus provides a framework for understanding the fundamental mechanisms acting on asteroids disrupted by natural impact[1,6].**

HST observed the ejecta once every 1.6 hours during the first 8 hours after DART's impact (Extended Data Table 1) at a viewing geometry shown in Fig. 1. The image collected at about T+0.4 hour (Fig. 2a) shows diffuse ejecta with several linear structures and clumps (concentration of materials ejected at similar velocities) spanning nearly the entire eastern hemisphere of Didymos. After ~T+2 hours, the initial, diffuse dust cloud had mostly dissipated, and an overall cone-shaped ejecta morphology emerged with the edges of the hollow cone manifested as two linear features (l7 and l8) due to the optical depth effect. The ejecta cone exhibited many distinct morphological features (Fig. 2b – 2f), some visible in multiple images over 3 to 10 hours and extending to nearly 500 km from the asteroid. These features moved radially away from the asteroid at constant speeds between a few and ~30 m/s as projected in the sky (Extended Data Table 2). Their radial expansion motion suggests that this material is directly ejected out of the Didymos system without being appreciably influenced by the gravity of the system or by solar radiation pressure. Based on the position angles (angle measured from north toward east) of the cone and a simple model (Methods), we find that the observed ejecta cone is consistent with a 3D opening angle of 125°±10° and centerline at a position angle of 67±8°, almost parallel to the incoming direction of the DART spacecraft. The observed ejecta cone is wider than the ejecta produced by vertical impact cratering experiments on granular media[7,8], though wider opening angles could be explained by the curvature of target surface[9] and the angle of internal friction of the target[10], as well as the geometry of the projectile[11].

Dimorphos's ejecta was distinctive from the ejecta of Comet 9P/Tempel 1 produced by Deep Impact[12], the only previous planetary impact experiment of comparable scale (Extended Data Figure 1a – 1c). Both experiments delivered similar momentum to their targets; the Deep Impact spacecraft carried 80% more kinetic energy than the DART spacecraft but the 6-km diameter nucleus of Tempel 1[12] was considerably more massive than the 151-m Dimorphos[2]. At the scale of HST, the Deep Impact ejecta were diffuse and mostly featureless, expanding at an average speed of ~100 m/s and a maximum speed of ~300 m/s[13,14]. This difference in ejecta morphology is likely due to the different target compositions and subsurface structures. While Tempel 1 has a highly porous subsurface[15] composed of fine-grained dust and rich in volatiles[16,17], the bouldery surface and potential rubble-pile interior of Dimorphos[2] could perturb the ejecta curtain and produce inhomogeneous structures in the ejecta[18,19].

From ~T+0.7 to 2.1 days, the ejecta features composed of slower dust escaping at <~1 m/s emerged from the base of the ejecta cone (Fig. 3a – 3d). The ejecta during this stage is characterized by the curved ejecta streams in the north (s1) and south (s2), some small curvilinear features (l16 – l19) between them, and the slight wrapping of these features around Didymos. The gravity of Didymos, which accounts for 88% of the gravitational potential of the binary system at the impact site, slowly distorted the shape of the original ejecta cone and created different morphologies for s1 and s2. The dust ejected in the original northern cone edge (l7) was in close proximity to Didymos (Fig. 1). As suggested by numerical simulation predictions[20,21], this dust was accelerated by Didymos and the trajectories were bent before escaping the binary system, forming the northern curved stream s1 (Extended Data Figure 2). The end of s1 near the asteroid contains relatively slow particles, whose trajectories were bent more than those of the relatively fast particles further away, causing the near end to shift clockwise about Didymos resulting in an 18º twist. In contrast, the majority of the dust in the original southern cone edge (l8, Fig. 2) was launched away from Didymos. Thus, these trajectories are less affected by Didymos's gravity, leading to a less curved southern stream (s2) with its near end slowly wrapping around the asteroid over time (Fig. 3a – 3f). The small curvilinear features between the two streams (l16 – l19) were likely composed of dust ejected in the front or back side of the hollow ejecta cone, behaving more or less similarly to either of the two curved streams and slightly rotating from the original radial direction.

Beyond the gravitational influence of the Didymos system, solar radiation pressure naturally separates particles of different sizes along the sunward-antisunward direction because small particles are accelerated faster than large particles[22]. The northern stream (s1), situated roughly orthogonal to the sunward direction, was increasingly widened to form the observed wing-like shape, with a diffuse antisunward edge and a relatively sharp sunward edge (Fig. 3f – 3j). This sharp edge indicates a cutoff in the largest particle size in the ejecta. Because the southern stream was nearly aligned toward the sun, those particles were first slowed by solar radiation pressure before eventually being turned towards the antisunward direction. Starting from T+4.7 days, the particles moving at different speeds and directions in s2 due to the inhomogeneous distributions of dust in the ejecta were separated into individual features (l20 – l24, Fig. 3f). They reached maximum projected sunward distances of up to 150 – 200 km. All these individual features (l20 – l24) and the small curvilinear features (l16 – l18) between the two main streams were stretched along the sunward-antisunward direction with time by solar radiation pressure (Fig. 3f – 3i). The finer particles in feature l16 – l18, which were located to the north of Didymos, were pushed further and caught up to the larger particles ejected into s1 earlier, appearing to overlap with the wing-like structure and creating a more complex pattern (Fig. 3g – 3h).

As a result of solar radiation pressure, a dust tail started to emerge antisunward nearly opposite the ejecta cone at ~T+3 hours. This tail quickly stretched out to >1500 km projected length and exceeded the spatial coverage of our images (Fig. 4). Around T+5.7 days, the narrow tail showed a relatively bright and sharp southern edge and a parallel but more diffuse northern edge (Fig. 4h). The overall morphology of Dimorphos's tail is similar to that of P/2010 A2, an active asteroid likely triggered by impact[4,23,24] (Extended Data Fig. 1d, 1e). The ~1" width of the tail is consistent with an initial speed of the dust comparable to the orbital speed of Dimorphos, suggesting that the tail contains the slowest ejecta particles. Additionally, the early tail within T+2 days slightly curved towards the south (Fig. 4d, 4e), whereas after T+8 days the tail became slightly more fan-shaped (Fig. 4i – 4k). With radiation pressure sorting out particle size along the tail, the earliest tail at ~T+3 hours was dominated by µm-sized particles, whereas cm-sized particles dominated the portion of the tail inside the HST field of view in the final image. The brightness profile of the tail is related to the particle size distribution in the ejecta. Assuming a power law for the differential size distribution, we derived an exponent of -2.7±0.2 for particles of 1 µm – a few mm radius, and an exponent of -3.7±0.2 for larger particles up to a few cm (Extended Data Fig. 3). Ejecta particles were observed to continuously leave the Didymos system through the final images acquired after T+15 days (Extended Data Figs. 4, 5).

Additionally, a secondary tail appeared between T+5.7 and T+8.8 days (Fig. 4i – 4k) but was no longer discernible on T+18.5 days (Fig. 4l). It originated from the Didymos system and pointed about 4º further north of the original tail, creating an overall fan-shaped tail morphology during this timeframe. The cause of the secondary tail is unclear, and multiple mechanisms are to be explored (Methods, Extended Data Figs. 4, 6), though the morphologies are consistent with the previous observations of active asteroids that displayed multiple tails[25-28]. The whole evolutionary sequence of Dimorphos's ejecta discussed above is displayed in Supplementary Video.

The DART mission demonstrated definitively that impacts can activate asteroids, consistent with prior asteroid observations[1]. Our observations provided a basis for reassessing the previous observations of active asteroids thought to be triggered by impact. The evolution of Dimorphos's ejecta suggests that the observed particle size in active asteroid tails could depend on the age of the tail, consistent with the range of particle sizes measured in the multiple tails of active asteroids 311P/PanSTARRS[26]. The lack of sub-mm sized dust in the tail of P/2010 A2[4,5,24], therefore, could be a result of the observations occurring 10 months after impact. DART, as a controlled, planetary-scale impact experiment, provides a detailed characterization of the target, the ejecta morphology, and the entire ejecta evolution process. DART will continue to be the model for studies of newly discovered asteroids that show activity caused by natural impacts.

**Figure Legends**

Figure 1. Geometry of the Didymos system at the time of impact as viewed from HST. Sky north is in the up direction and east is on the left in this view. The equivalent diameters of Didymos (large spheroid) and Dimorphos (small spheroid) are 761 m and 151 m, respectively2. The orbit of Dimorphos around Didymos before the impact, depicted by the black circle, has a semimajor axis of 1.206±0.035 km$^3$ and an eccentricity <0.03$^{29}$. The sizes of Didymos and Dimorphos and their separation in the figure are to scale. The entire system lies within one pixel in our HST images. Dimorphos orbits clockwise with an orbital speed of ~0.17 m/s. The positive pole of Didymos (also the orbital pole of the system) is represented by the blue line, pointing close to the south celestial pole and 51º out of the sky plane away from the Earth. The Sun is at a position angle of 118º, represented by the orange line and the dot-circle symbol. The DART spacecraft vector is represented by the red line, with arrows, going from east to west at a position angle of 68º and within 1º of the sky plane.

Figure 2. Evolution of Dimorphos ejecta from T+0.4 to T+8.2 hours. All images are displayed in duplicate pairs, with the left unannotated for clarity, and the right having features marked by white markers and labels. The inset in the left of each panel is the 100-pixel wide region centered at the asteroids but with the flux scaled down by 10✕ to show the details of the bright core. "'x" marks artifacts due to residual cosmic rays, frame boundaries, background objects, defective pixels, etc.

The times correspond to the mid-observation time of each image. Black lines mark diffraction spikes. All images are displayed with the same logarithmic brightness stretch. Sky north is up and east to the left. The scale bars mark 200 km at the distance of Didymos. The yellow arrows point to the direction of the Sun, the cyan arrows the heliocentric velocity direction of Didymos, and the red arrows the direction of DART spacecraft at impact, all projected in the sky plane at the times of observations. HST had a pointing drift during the exposures of some images, causing smear of about 4 – 7 pixels in the first four images (before T+5.0 hour), and about 14 pixels in the T+6.6 hours image, all along the northeast-southwest direction (Methods). The drift widens the tail and the two diffraction spikes orthogonal to the direction of the drift. Most features are much larger than the length of the drift, and we added uncertainties to account for the effect of this drift in our measurements. Many features are visible, including linear features (l1 – l12), an arc (arc1), a circular feature (c1), blobs (b1 – b3), and a tail. The ejecta cone is marked by linear features l7 and l8.

Figure 3. Evolution of ejecta from T+0.7 days (T+17.8 hours), following Fig. 2, through T+18.5 days. The inset, image orientation, brightness stretch, scale bars, and vector arrows are all the same as in Fig. 2. The main characteristics of the ejecta during this period of time include the curved ejecta streams (s1 and s2), linear features (l7, l11 – l24), blobs (b3 – b5), a circular feature (c1), and an arc (arc2). The original north edge of the ejecta cone (l7) is still visible in images before T+5.7 days (panels a – g). The early southern curved stream (s2) could be overlapped with the southern edge of the original ejecta cone (panels a – e), which is not separately marked. The northern curved stream (s1) is widened along the tail direction in about T+5 days, forming a wing-like feature (panels g – k). A group of linear features (l16 – l24), some being part of the southern curved stream (l21 – l24), showed a clockwise rotation about Didymos from T+1.1 to T+4.7 days (panels b – f). These linear features later (T+5.7 days) stretched along the tail direction under solar radiation pressure (panel g – i), with those in the north of Didymos overlapping with the wing-shaped feature. A secondary tail is visible between T+8.8 and T+14.9 days (panels h – j, also see Fig. 4). The curved edge of the wing-like feature is visible in the last image (panel k).

Figure 4. Tail formation from the Dimorphos ejecta cloud. All frames are rotated such that the expected direction of the tail based on our dust dynamic model (Methods) is in the horizontal direction extending towards the right. All frames are displayed in the same logarithmic brightness scale. The regions outside the field of view are marked by a dark blue color. The scale bars are aligned with the asteroid on one end and extend 200 km long towards the tail direction. Note that the first three frames (a, b, c) have pointing-induced drift in the plane of sky of 5 – 7 pixels approximately along the direction of the vertical diffraction spikes. The drift in all other frames is < 2 pixels. The first frame (a) in this sequence acquired at T+0.08 days (T+1.9 hours) shows no signs of a tail. A tail was visible starting from the second frame (b) acquired at T+0.15 days (T+3.5 hours). The tail continued to grow in a direction that is in general consistent with an impulsive emission of dust from Dimorphos at the time of impact. The secondary tail is visible between T+8.82 and T+14.91 days (panels i – k), pointing at about 4º north of the original tail.

# Methods

1. **Observations and data reduction and processing**

We used a total of 19 HST orbits (period 95 min) over about 19 days to observe the Dimorphos ejecta (Extended Data Table 1). The first orbit (orbit 0o) was before impact. The second orbit through the 7th orbit (orbits 01 – 06) started about T+15 min, and continuously observed the ejecta except for during Earth occultations of the target. In the next 5 orbits (orbits 11 – 15), we observed the ejecta roughly once every 12 hours, and then once every day in the following three orbits (orbits 16 – 18). In the final phase (orbits 21 – 24) observations were executed once every 3 days. The observations concluded 18.5 days after impact. In each orbit, images were collected at multiple exposure levels, where the central core of Didymos was unsaturated in short exposures, and long exposures saturated Didymos to image the relatively faint ejecta and tail. All images were collected through filter F350LP (pivot wavelength 587 nm, bandwidth 149 nm)[30].

The observations were planned to track at the Dimorphos ephemeris rate. The tracking nominally included corrections for parallax due to HST's orbit around the Earth and was expected to keep Didymos inside the field of view with minimal drift in the field of view for all exposures. However, due to an as-yet unexplained tracking problem, some orbits lost the target in various numbers of exposures, and some long exposures included pointing drift of more than 10 pixels. We limited our analysis to those exposures with less than 7 pixels of drift, and occasionally used long exposures with more drift when no good images were available for the particular orbits.

Images were calibrated by the HST standard calibration pipeline at the Space Telescope Science Institute[31]. We then removed the sky background measured from a square 100 – 400 pixels wide and 100 – 300 pixels from the top right corner, depending on the image size. This area is in general 20" away from Didymos and shows no sign of any ejecta.

Aperture photometry was measured in all short, unsaturated exposures that have been corrected for charge transfer efficiency (CTE)[31] but not geometric distortion (_flc files). The centroid was defined by a 2D Gaussian fit with a 5×5 pixel box centered at the photocenter. The pixel area map was used to correct pixel area variations in the image[31]. The total counts were measured with circular apertures of 1 – 130 pixels radius (0.04" - 5.2"). We converted the total counts to flux density and Vega magnitude based on the photometric calibration constants (PHOTFLAM = $5.3469 \times 10^{-20}$ erg / (Å $cm^2$ electron), PHOTZPT = 26.78) provided in the FITS headers and HST photometric calibration website. The total brightness of Didymos including the ejecta and the total brightness of ejecta are shown in Extended Data Fig. 4.

We used the CTE-corrected and geometric distortion corrected images (_drc files) to study the morphology of the ejecta. In order to increase the signal-to-noise ratio of the faint ejecta features, we stacked all long exposures in each orbit because no change is visible in the ejecta morphology within each orbit. The centroid of long exposures that are saturated in the center was determined by the cross-section of the diffraction spikes. Some long exposures with pointing-induced drift were included in the stack, but those with more than 10 pixels of drift were discarded.

The effects of such drift are accounted for as additional positional uncertainties to the measurements of features, which are mostly smaller than the length of the drift. Cosmic rays and background stars were rejected in the stacking process. Because different numbers of good long exposures were available in each orbit, the total exposure times vary from 25 s – 50 s in most stacked long exposures and reach 155 s for the orbit 21 stack and 110 s for the orbit 23 stack.

Various image enhancement techniques commonly used for studies of comets (see review by ref. 32) were used to assist the identification of ejecta features, including azimuthal median subtraction and division, re-projection to azimuthal and radial projection, and different brightness stretching and displaying with various color tables. All identified features were cross-confirmed by multiple techniques.

The speeds of features as projected in the image plane were estimated by assuming that all features originated from the asteroid at the time of impact and moved directly away from the asteroid. The projected distance of a feature from the asteroid and the corresponding observation time yielded the projected speed of the feature. Note that the speeds estimated this way do not represent the true terminal speeds of the features after escaping the binary system for slow ejecta (<~1 m/s), or for features affected by solar radiation pressure. Because in those case, their trajectories are appreciably affected by the gravity of Didymos (Extended Data Fig. 2) or solar radiation pressure.

## 2. Ejecta cone opening angle and direction

We based our ejecta cone characteristics on the ejecta structures moving at >1 m/s in the images within T+8.2 hours (Fig. 2). These structures showed a linear motion moving away from the asteroid along the radial direction from the binary asteroid (Extended Data Table 2). Assuming that the majority of the ejecta dust is within a thin cone-shaped curtain, the two edges of the cone would appear as two bright rays along the radial direction from Dimorphos due to the optical depth effect when viewed from the side. Because the DART impact velocity is close to the sky plane (Extended Data Table 1), if we assume that the cone direction is close to the inverse of the DART impact velocity direction, the cone is close to being viewed from the side in HST images, and the opening angle spanned by the two edges of the cone (linear feature l7 and l8) is close to its 3D opening angle (see below).

We measured the position angles of the two edges of the ejecta cone from both the original image and the enhanced images (see Method S1). The uncertainty range of the position angles is defined by the apparent width of the linear feature. Our measurement resulted in an ejecta cone centered within 5º of the incoming direction of DART with an opening angle of about 130º. Because of the fuzziness of the ejecta rays and their slight curvature, the uncertainty of the measured position angles could be as high as ±8º, resulting in an uncertainty of the opening angles up to ±12º. Taking the mean of these two edges and the maximum value of the uncertainty yields the ejecta cone axis at a position angle 67±8° under the assumption that the ejecta cone is axisymmetric along the cone axis.

To further constrain the ejecta cone geometry, we constructed a 3D numerical cone model parameterized by the direction of the cone axis in Right Ascension (RA) and Declination (Dec), as well as an opening angle, to compare with the images. We first projected the six early post-impact images (Fig. 2) in an azimuthal-radial projection and, for each image, generated a histogram of pixels brighter than 18 mag/arcsec² along the azimuthal direction. The azimuthal bins with the highest pixel counts (except for those of the tail and diffraction spikes) define the two cone edges with more or less a Gaussian distribution. The mean and the 1-σ uncertainty of the position angles of the two cone edges are derived from the histograms. On average, the northern and southern cone edges are at position angles of 4°±8° and 131°±8°, respectively, consistent with the measurements described above. We then generated simulated images from the model ejecta cone and computed the corresponding histogram following the same approach for the actual images. This histogram was compared to the measured cone edge position angles to calculate a score, defined as,

$$f = \sum_{i=1}^{2} \sum_{j=1}^{n} \frac{s_j}{\sigma_i \sqrt{2\pi}} exp\left(-\frac{1}{2} \frac{(x_j - \mu_i)^2}{\sigma_i^2}\right)$$

where $\sigma_i$ and $\mu_i$ are the standard deviation and mean of the northern or southern edge ($i$ = 1, 2), respectively, $x_j$ is the position angle of the histogram bin $j$ for the simulated image, and $s_j$ is the pixel count in bin $j$. We searched the cone axis in the full range of RA and Dec, and the opening angle in 100° – 160° for the highest score. Because HST images alone could not determine whether the cone faced toward or away from the Earth, this approach resulted in a pair of best-fit cone axis solutions that were symmetric with respect to the image plane. We thus considered both as feasible cone axis directions. The uncertainties of the solutions were estimated with 500 random samples of the measured cone edge position angles distributed in two Gaussians with the measured means and standard deviations. The best-fit cone axis directions were (RA, Dec) = (141°±8°, 25°±6°) and (120°±9°, 10°±7°), both with an opening angle of 125°±10° (1-σ uncertainties). Both solutions are about 12° from the image plane, with the former pointing toward the Earth and the latter pointing away.

### 3. Dynamic model of the tail

The position angle of the tail and its uncertainty were determined by the radial directions from the asteroid that define the visible boundary of the tail at the furthest point along the tail in all (short and long exposures) stacked images that contain the tail. The dust dynamics model under the influence of solar radiation pressure follows ref. 22, where the motion of dust is determined by $\beta_{srp}$, which is defined as the ratio of the solar radiation pressure force to the solar gravitational force. $\beta_{srp}$ depends on particle radius, $r$, and density, $\rho$, as

$$\beta_{srp} = KQ_{pr}/\rho r$$

where $K$ = 5.7x10⁻⁴ kg/m² is a constant, $Q_{pr}$ is the radiation pressure coefficient averaged over the solar spectrum, which is usually assumed to be 1. We assumed a grain density of 3.5x10³ kg/m³ for the dust in the ejecta, following the density of ordinary chondrite meteorites[33], considering that Didymos-Dimorphos system shows an S-type spectrum that is associated with (LL) ordinary chondrite material[34].

Pre-impact modeling suggested that the acceleration of solar radiation pressure always exceeds that of the gravitational acceleration of the Didymos system for ejecta particles < 100 μm in size[20,35]. These small particles are pushed out of the binary system in less than 10 hours. Didymos's gravity is predominant within about 3 km for mm particles, and 10 km for cm particles.

The modeling of the orientation of the tail in the sky plane follows the synchrone-syndyne approach[36], where synchrones are the loci of dust particles ejected with zero initial velocity at the same time but with various $\beta_{srp}$. The measured position angles of Dimorphos's tail coincide to within 4° of the direction suggested by the synchrones associated with the time of impact in all images, suggesting that solar radiation pressure dominates the tail formation (Extended Data Fig. 7). The small discrepancy between T+1 and T+5 days is likely due to the slight apparent curvature of the tail (Fig. 4e – 4h), which may be related to the non-zero average initial velocity of dust particles with respect to the binary system inherited from the orbital speed of Dimorphos.

The non-zero initial velocity of ejecta dust causes the tail to widen. The average initial velocity of Dimorphos's ejecta, as projected in the image plane, has a northward component, which causes the tail to widen toward the north with respect to the loci of the hypothetical zero-velocity particle (synchrone). The relatively sharp southern edge and the more diffuse northern edge are consistent with the expectation from the ejecta mass-speed relationship[37] because the number of dust particles decreases with increasing ejection speeds. The 1" width of the tail is consistent with an initial velocity dispersion $\Delta v$ = 0.15 m/s, comparable to the orbital speed of Dimorphos, suggesting that the tail is primarily composed of the slowest ejecta.

The inverse proportionality of $\beta_{srp}$ with particle size means that small particles experience stronger solar radiation pressure and are pushed away from the asteroid faster after ejection than large particles. Because the duration of our HST observations is much shorter than the orbital period of Didymos around the Sun (2.1 years), the motion of particles along the tail relative to the asteroid under solar radiation pressure can be approximated by a constant acceleration motion. As the length of the tail grows, particles of various sizes spread out along the tail, with the smallest particles remaining near the far end of the tail from the asteroid, while larger particles dominate the end near the asteroid. Assuming a power-law differential particle size distribution with an exponent of $\alpha$ for the tail, we derived that the brightness of the tail is expected to have a power-law relationship with the distance to the asteroid with an exponent $b$ = -4 - $\alpha$.

We extracted the brightness profiles of the tail from stacked long exposures from T+5 hours until the last stack at T+18.5 days (Extended Data Fig. 3). The exponent $\alpha$ of the differential particle size distribution was derived from the linear part of the tail brightness profiles (in log-log space) in various images, corresponding to a range of $\beta_{srp}$ from 0.2 to 8x10$^{-4}$, to be within -2.2 and -3.1 with an average of -2.7 and a standard deviation of 0.2. The range of $\beta_{srp}$ indicates particle sizes between about 1 μm and a few mm. In images after about T+6 days, the tail brightness displays two regions with different power law slopes. The inner region appears to be influenced by the particles in the curved ejecta streams that started to overlap with the tail. The outer region has best-fit slopes close to -2.7 as in the early images, whereas the slope of the inner region ranges from -3.6 to -3.9. The range of $\beta_{srp}$ for the inner region is $7\times10^{-4}$ to $1\times10^{-5}$, corresponding to the large mm – cm sized particles. The lack of small particles in the curved streams is expected because 100 µm or smaller particles should have been removed a few hours after impact. The apparent increasing steepness of the particle size distribution in this size range also seems to indicate that the bulk of ejecta particles have a size cutoff at a few cm. If the particle size distribution of the tail represents that of the ejecta, then a power law index of -2.7 means the total ejecta mass is dominated by the largest particles.

The above treatment assumes that the albedo is independent of particle size, which needs to be examined. Based on laboratory phase function measurements of micron-size aerosols[38] and mm-size particles[39], along with supporting models of scattering efficiency[40], the albedo of µm-size particles is about 70% that of mm-grains at the phase angle of our early observations (54º). This brightness ratio is reversed at the phase angle corresponding to the final images (74º), where µm-size particles become about 16% brighter. Our calculation suggests that the small difference between the albedos of µm- and mm-sized particles changes the best-fit power law index of the particle size distribution by less than 2%. Our assumption of the same albedo throughout the µm- to cm sized particles holds.

## 4. Secondary tail

The small decrease of the overall fading rate of the Didymos system total brightness between about T+5 and T+7 days indicates an increase in the total scattering cross-section in the ejecta within 10 km of the system (Extended Data Fig. 4), partly compensating for the ejecta moving out of the photometric aperture. It is unlikely to be caused by albedo change for the ejecta particles. Injection of new dust particles into the ejecta is considered.

This scenario and its timing are also supported by the synchrone model (Extended Data Fig. 6), where the projected direction of the secondary tail is consistent with the synchrones associated with about T+5.0 to T+7.1 days. The similar narrow width of the secondary tail with the original tail suggests a low initial velocity of ~0.15 m/s for the dust particles. While the Didymos binary environment could complicate the dust motion and cause deviation from the zero initial velocity assumption of the idealized synchrone model, the observed low initial velocity of the dust in the secondary tail implies limited effects.

The possible mechanisms of the secondary dust emission could include the re-impact of ejecta blocks onto Dimorphos or Didymos[35], or large ejecta blocks disintegrating into small pieces due to spin up or mutual collisions[41]. Mass shedding from the surface of Dimorphos due to rotation is not likely given its slow rotation if its spin is tidally locked. But mass movement and shedding from Didymos could potentially be triggered by ejecta re-impact due to its fast rotation causing a net outward acceleration at its equator[42], though no clear indication has been confirmed yet[3].

Once dust is lifted from the surface of Dimorphos or Didymos via these mechanisms, solar radiation pressure will quickly sweep the dust into the antisolar direction, forming a secondary tail.

Other mechanisms, such as the dynamic interaction between the slow ejecta dust and the binary system[43], gravitational scattering for the ejecta dust when they are turned back by solar radiation pressure and pass the binary system, or photon-charged dust particles under the influence of interplanetary magnetic field[44] could also result in unusual tail morphology leading to the appearance of a secondary tail. Our dynamic simulations suggested that a secondary dust emission is not necessary to form a secondary tail that has consistent morphology as the observed. But these scenarios may not be accompanied by the increase of ejecta dust as suggested by the fading lightcurve of the Didymos system.

**Methods References**

**Data availability** All raw HST data associated with this article are archived and are publicly available at the Mikulski Archive for Space Telescopes (https://mast.stsci.edu/search/ui/#/hst/results?proposal_id=16674) hosted by the Space Telescope Science Institute. The stacked long exposures presented in Figures 2 – 4 are available from a website hosted at JHU/APL (https://lib.jhuapl.edu/papers/ejecta-from-the-dart-produced-active-asteroid-dimo). Other related data are available upon request.

**Acknowledgments** This work was supported by the DART mission, NASA Contract No. 80MSFC20D0004 and by the Italian Space Agency (ASI) within the LICIACube project (ASI-INAF agreement AC n. 2019-31-HH.0). Part of this research was carried out at the Jet Propulsion Laboratory, California Institute of Technology, under a contract with the National Aeronautics and Space Administration. J.-Y.L. acknowledges the support provided by NASA through grant HST-GO-16674 from the Space Telescope Science Institute, which is operated by the Association of Universities for Research in Astronomy, Inc., under NASA contract NAS 5-26555. L.K. acknowledges support from NASA DART Participating Scientist Program, Grant #80NSSC21K1131. R.L., D.A.G., and T.J.S. acknowledge funding from the NASA/GSFC Internal Scientist Funding Model (ISFM) Exospheres, Ionospheres, Magnetospheres Modeling (EIMM) team, the NASA Solar System Exploration Research Virtual Institute (SSERVI), and NASA award number 80GSFC21M0002. R.M. acknowledges support from a NASA Space Technology Graduate Research Opportunities (NSTGRO) Award (Contract No. 80NSSC22K1173). P.M. acknowledges funding support from the European Union's Horizon 2020 research and innovation program under grant agreement No. 870377 (project NEO-MAPP), the CNRS through the MITI interdisciplinary programs, CNES and ESA. F.F. acknowledges funding from the Swiss National Science Foundation (SNSF) Ambizione grant No. 193346. J.O. has been funded by grant No. PID2021-125883NB-C22 by the Spanish Ministry of Science and Innovation/State Agency of Research MCIN/AEI/ 10.13039/501100011033 and by "ERDF A way of making Europe". G.T. acknowledge financial support from project FCE-1-2019-1-156451 of the Agencia Nacional de Investigación e Innovación ANII (Uruguay). T.K. is supported by Academy of Finland project 335595 and by institutional support RVO 67985831 of the Institute of Geology of the Czech Academy of Sciences. F.M. acknowledges financial support from grants CEX2021-001131-S funded by MCIN/AEI/10.13039/501100011033 and PID2021-123370OB-I00. Research by M.G.

is supported, in part, by the Academy of Finland grant 345115. The authors thank Joseph DePasquale (STScI) for generating the animation included in the Extended Data.

**Author Contributions** J.-Y.L. is the PI of the HST program (GO-16674), together with Co-Is M.M.K., C.A.T., A.S.R., S.C., L.K., A.F.C., E.G.F., to observe the DART ejecta, and leads the effort to develop this paper. M.H., T.L.F., M.M.K. contributed to the ejecta cone measurement and modeling and ejecta evolution study. G.T. contributed in the photometry, formation of the tail, and the comparison with active asteroids. F.M., A.C.B., B.M., C.O., J.-B.V. contributed in the study of the formation of the tail. S.C. contributed to photometric study of the ejecta. J.M.S., S.D.R, M.J., C.M.E., and A.M.S. contributed to the understanding of ejecta features related to impact. L.D. and A.V. supported the scheduling, review, and testing of the observing sequence. F.F., S. Ivanovski, A.R., D.J.S., S.S. contributed to dynamical modeling of the ejecta. R.L., D.A.G., T.J.S. supported the derivation of dust size distribution from dust light scattering properties. A.F.C. and A.S.R. are the DART Investigation Team Lead. N.C. is the DART Coordination Lead. C.A.T. is the lead of the DART Observations Working Group, providing general observations support. E.G.F. is the lead of DART Ejecta Working Group, providing support to the interpretations and modeling of ejecta. N.A.M. supported C.A.T to provide general observations support. S.B., M.G. contribute to the ejecta particle size study. M.T.B., G.C., E.D., R.T.D., E.M.E., M.I.H., P.H.H., S. Ieva, S.J., T.K., A.L., T.L., Z.-Y.L., P.M., R.M., J.O., M.P., C.S., J.S., P.S., S.R.S., J.M.T.-R., A.F., T.K, T.K, A.M., L.K., F.L.F., M.L., H.W., provided comments and improvements to the manuscript.

**Competing interests** The authors declare no competing interests.

**Additional information** Correspondence and requests for materials should be addressed to Dr. Jian-Yang Li (jyli@psi.edu).

**Extended Online Data Legends**

Extended Data Table 1. HST observations of DART impact. This table lists the times and observing geometries for each of the 19 orbits of HST observations. The parameters listed in this table only refer to usable images in each orbit. Some images were lost due to tracking problems.

Extended Data Table 2. Selected features and their approximate plane of sky speed. The speeds reported here are averaged over all measurements in multiple images for every feature. The scatters in measured speeds are typically < 5% for each feature. The measurement is based on the approximate distance of the feature to the asteroid and the corresponding mid-observation times.

Extended Data Figure 1. Comparison of the ejecta and tail morphology of Dimorphos with other objects. (a) Deep Impact ejecta approximately one hour after impact observed by HST[13]. (b)

Dimorphos ejecta approximately T+0.4 hour (Fig. 2a). (c) Dimorphos ejecta approximately T+5 hours (Fig. 2d). (d) Tail of P/2010 A2 observed by HST on January 29, 2010 at a distance of 1.09 au[4] (original image by NASA, ESA, D. Jewitt (UCLA), source: https://hubblesite.org/contents/media/images/2010/07/2693-Image.html?news=true, rotated to approximate north up). (e) Dimorphos tail observed on T+5.7 days (Fig. 4h). All images are displayed with north in the up direction and east to the left.

Extended Data Figure 2. Illustration of curved ejecta streams seen by HST on T+2.1 days (Fig. 3d). (a) The red lines represent the trajectories of eight dust particles ejected at 0.43 m/s, each involved in the northern or southern edges of the ejecta cone. The initial directions are based on the measured cone geometry (Methods). The trajectories are curved by the gravity of Didymos and Dimorphos. The curved dark blue lines are the locations of multiple particles ejected at different speeds along the same direction as the particle in each corresponding red curve, forming the observed curved ejecta streams. The area in the illustration is 600 km wide. (b) Same illustration as (a) but with a smaller scale, showing the more remarkable curvature in the ejecta streams near the binary system. These streams capture a snapshot of particles' positions with initial ejection speeds less than <~ 1 m/s.

Extended Data Figure 3. Tail brightness profile and ejecta particle size distribution. (a) Brightness profiles along the tail from various images. The dashed lines are average surface brightness extracted along the tail with a width of 40 pixels (1.6"), offset vertically for clarity. The solid lines are corresponding best-fit power law models. Two sections are fitted separately for the profiles from the images collected on and after October 2, as described in the text. (b) Best-fit power law index for the differential size distribution (dSFD) of ejecta dust particles with respect to $\beta_{srp}$ on the bottom axis and the corresponding particle radius (assuming a density of 3500 kg/m$^3$) on the top axis. Filled circles are derived from the main tail, open triangles from the secondary tail. The horizontal error bars represent the range of $\beta_{srp}$ covered by the corresponding tail profile. The colors of symbols correspond to the colors of profiles in panel (a). The slope values from the outer section have $\beta_{srp}$ higher than 1x10$^{-4}$, and those from the inner section correspond to $\beta_{srp}$ between 1x10$^{-4}$ and 1x10$^{-5}$. The dashed horizontal line is the average -2.7 for the outer sections, and the green shaded area represents the standard deviation.

Extended Data Figure 4. Brightness evolutions of Didymos and the ejecta. (a) Total magnitude of Didymos in 10 km, 30 km, and 50 km radius apertures at the distance of Didymos measured from HST images as a function of time after impact. (b) Magnitude of ejecta with respect to time after impact. The black curve in both panels is the magnitude of Didymos based on the IAU HG phase function model with a G=0.20[45], scaled to match the observed pre-impact magnitude. The ejecta magnitude corresponds to the difference between the observed total flux and the flux from Didymos. The ejecta is brighter than Didymos for about 2.5 days after impact in the 10 kmradius aperture.

Extended Data Figure 5. Azimuthally averaged radial profiles of Didymos and ejecta. The curves are extracted from the pre-impact image (-0.1 d) and the last three images (+11.9d, +14.9d, and +18.5d). The widened PSF profiles of late images suggest a slightly extended source due to ejecta

dust close to the asteroid. 1 pixel corresponds to 0.04" or 2.1 – 2.3 km at the distance of Didymos in the last three images.

Extended Data Figure 6. Synchrone analysis of the main tail and the secondary tail. (a) Image taken at T+11.86 days is displayed in logarithmic brightness stretch. North is up and east to the left. The features marked by "x" are artifacts from a background object and a cosmic ray hit. (b) Same image as in (a) but with synchrones corresponding to various dates overlaid. The direction of the main tail is consistent with the synchrone at impact time (T+0.0 days), and the secondary tail is consistent with the synchrones between T+5.0 and T+7.1 days.

Extended Data Figure 7. The position angles of the tail measured from HST images. The blue circles are measured from the stacked images of the short exposures, and the orange circles are measured from the stacked images from the long exposures. The green triangles are the position angles of the secondary tail. The red dashed line is the antisolar direction, and the blue solid line is the position angle of synchrones for dust emitted at the time of impact. The tail orientation measured from the short exposures could be affected by the secondary tail due to the low signal-to-noise compared to the long exposures.

**Figure 1.**

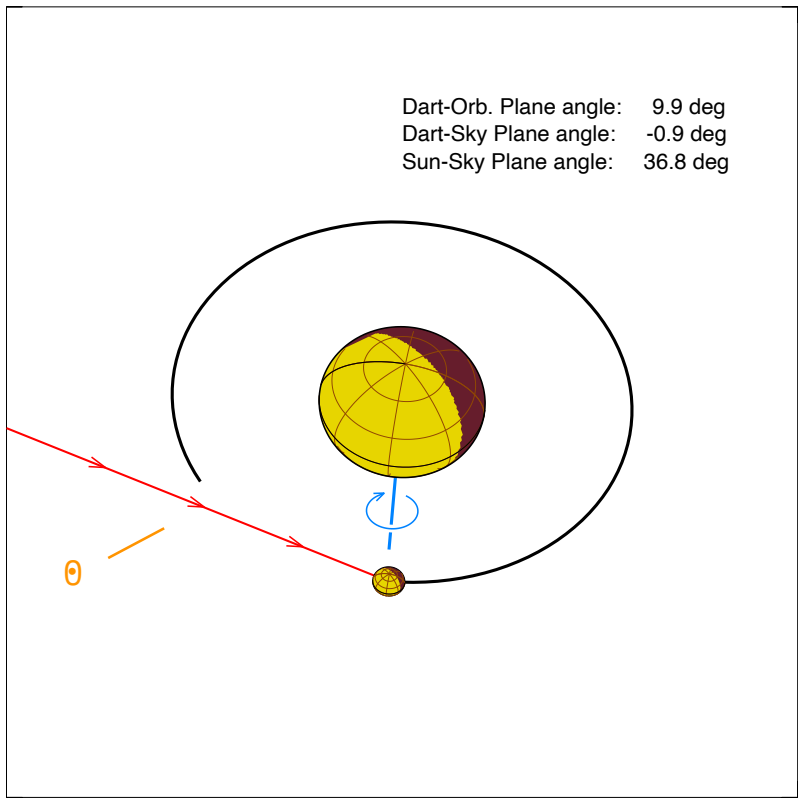

**Figure 2.**

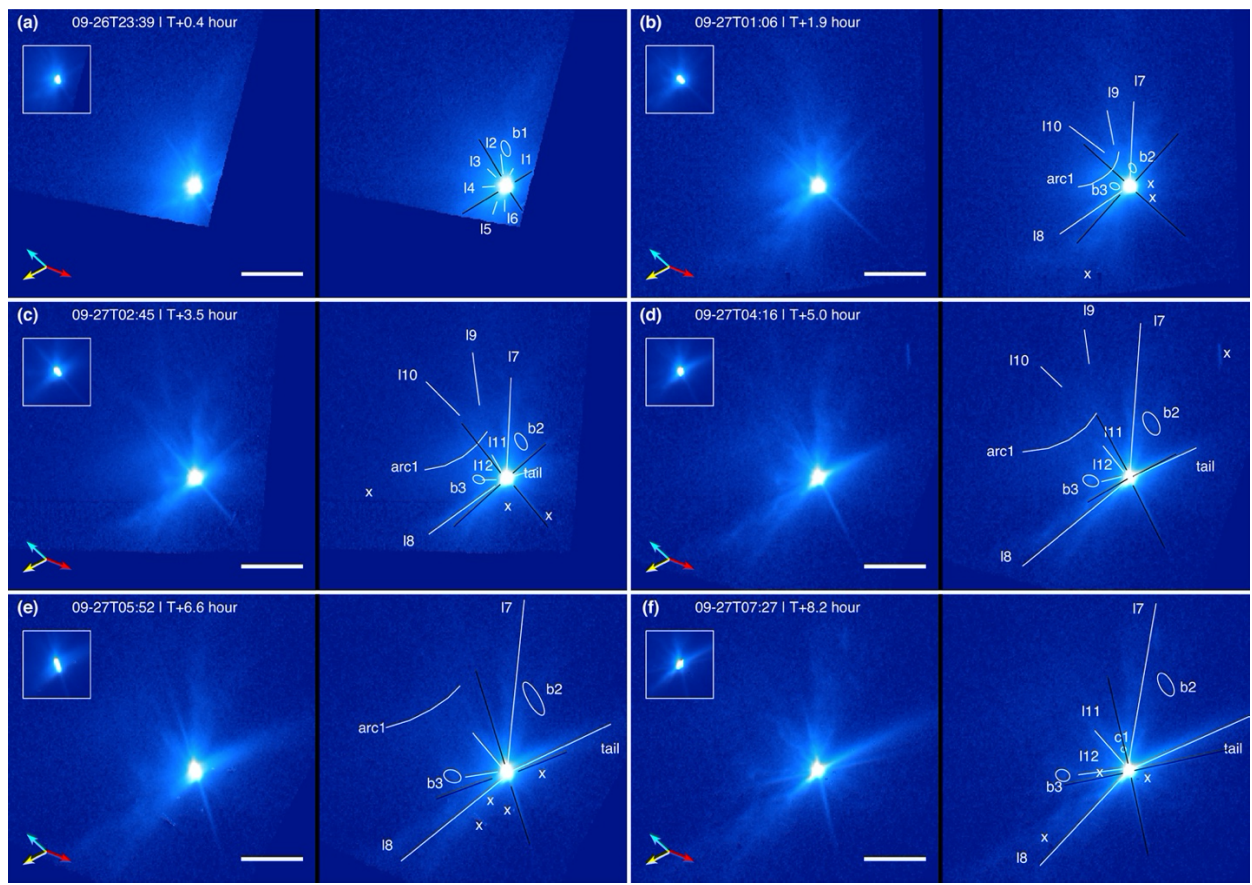

**Figure 3.**

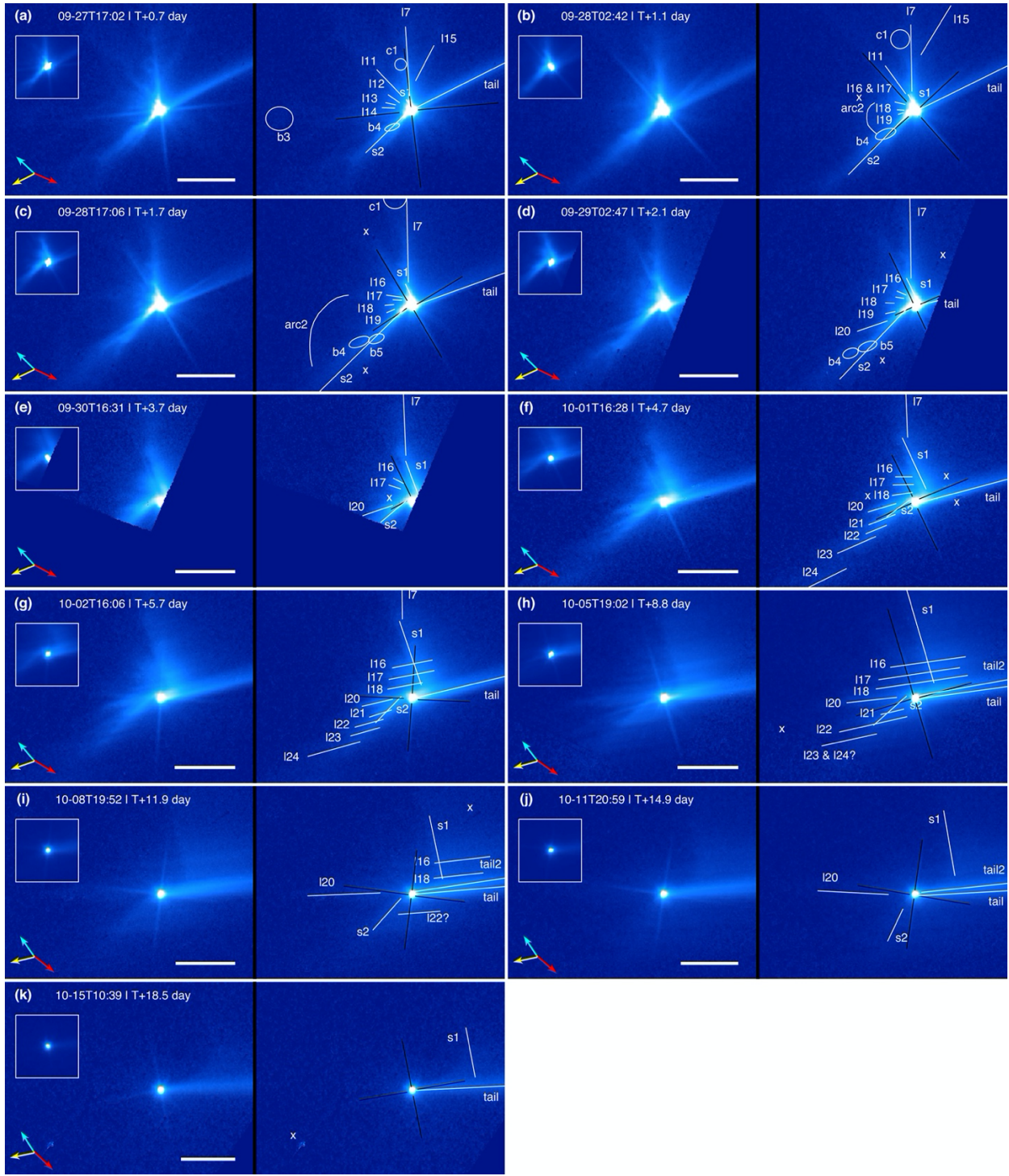

**Figure 4.**

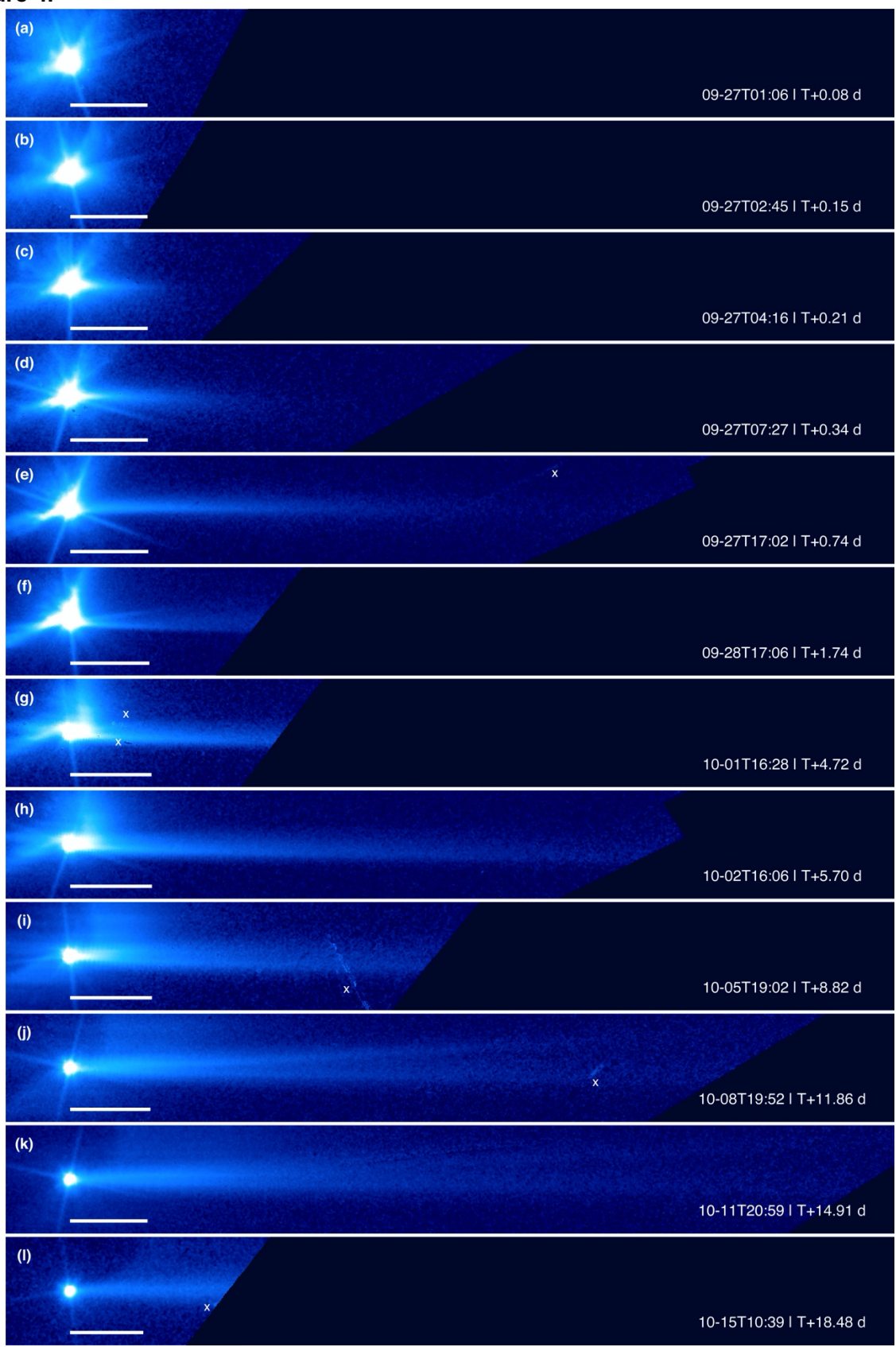

**Extended Data Table 1.**

| Orbit ID | Start and End UTC (Year 2022) | Time from Impact | Geocentric Distance (au) | Phase Angle (deg) | DART Velocity Position Angle (deg) | DART Velocity Tilt (behind image plane, deg) |
|---|---|---|---|---|---|---|
| 0o | 09-26T21:54 – 09-26T21:59 | -1.35 – -1.26 hour | 0.0757 | 53.2 | 68.2 | 0.76 |
| 01 | 09-26T23:29 – 09-26T23:39 | 0.24 – 0.40 hour | 0.0756 | 53.3 | 68.1 | 0.94 |
| 02 | 09-27T01:04 – 09-27T01:12 | 1.82 – 1.96 hour | 0.0755 | 53.4 | 68.0 | 1.1 |
| 03 | 09-27T02:40 – 09-27T02:50 | 3.41 – 3.58 hour | 0.0755 | 53.5 | 67.9 | 1.3 |
| 04 | 09-27T04:15 – 09-27T04:27 | 5.00 – 5.20 hour | 0.0754 | 53.6 | 67.8 | 1.4 |
| 05 | 09-27T05:50 – 09-27T05:58 | 6.59 – 6.72 hour | 0.0753 | 53.7 | 67.7 | 1.6 |
| 06 | 09-27T07:25 – 09-27T07:44 | 8.17 – 8.50 hour | 0.0752 | 53.8 | 67.6 | 1.8 |
| 11 | 09-27T16:57 – 09-27T17:31 | 17.7 – 18.3 hour | 0.0748 | 54.3 | 66.9 | 2.8 |
| 12 | 09-28T02:28 – 09-28T03:02 | 1.13 – 1.16 day | 0.0744 | 54.9 | 66.2 | 3.9 |
| 13 | 09-28T16:45 – 09-28T17:20 | 1.73 – 1.75 day | 0.0738 | 55.8 | 65.3 | 5.4 |
| 14 | 09-29T02:17 – 09-28T02:52 | 2.13 – 2.15 day | 0.0735 | 56.4 | 64.6 | 6.5 |
| 15 | 09-29T16:34 – 09-29T16:38 | 2.72 – 2.72 day | 0.0730 | 57.3 | 63.7 | 8.0 |
| 16 | 09-30T16:23 – 09-30T16:47 | 3.71 – 3.73 day | 0.0723 | 58.8 | 62.1 | 10.7 |
| 17 | 10-01T16:12 – 10-01T16:47 | 4.71 – 4.73 day | 0.0718 | 60.3 | 60.6 | 13.5 |
| 18 | 10-02T16:01 – 10-02T16:35 | 5.70 – 5.72 day | 0.0714 | 61.7 | 59.1 | 16.2 |
| 21 | 10-05T18:38 – 10-05T19:12 | 8.81 – 8.83 day | 0.0713 | 66.0 | 55.0 | 24.8 |
| 22 | 10-08T19:40 – 10-08T20:15 | 11.85 – 11.87 day | 0.0727 | 70.0 | 51.7 | 33.0 |
| 23 | 10-11T20:42 – 10-11T21:16 | 14.89 – 14.92 day | 0.0753 | 72.4 | 49.1 | 40.7 |
| 24 | 10-15T10:26 – 10-15T10:40 | 18.47 – 18.48 day | 0.0797 | 74.8 | 46.8 | 48.8 |

**Extended Data Table 2.**

| Feature | Visible in Figure/Panel | Approximate Apparent Speed (m/s) |
| --- | --- | --- |
| b1 | Fig. 2a | 81 |
| b2 | Fig. 2b, c, d, e, f | 11 |
| b3 | Fig. 2b, c, d, e, f, Fig. 3a | 7.1 |
| arc1 | Fig. 2b, c, d, e | 12 – 21 |
| l9 | Fig. 2b, c, d | 21 – 33 |
| l10 | Fig. 2b, c, d | 20 – 33 |
| l15 | Fig. 2a, b | 2.9 |
| c1 | Fig. 2f, Fig. 3a, b, c | 2.6 |
| b4 | Fig. 3a, b, c, d | 1.4 |
| b5 | Fig. 3c, d | 1.1 |
| arc2 | Fig. 3b, c | 1.4 – 2.1 |

**Extended Data Figure 1.**

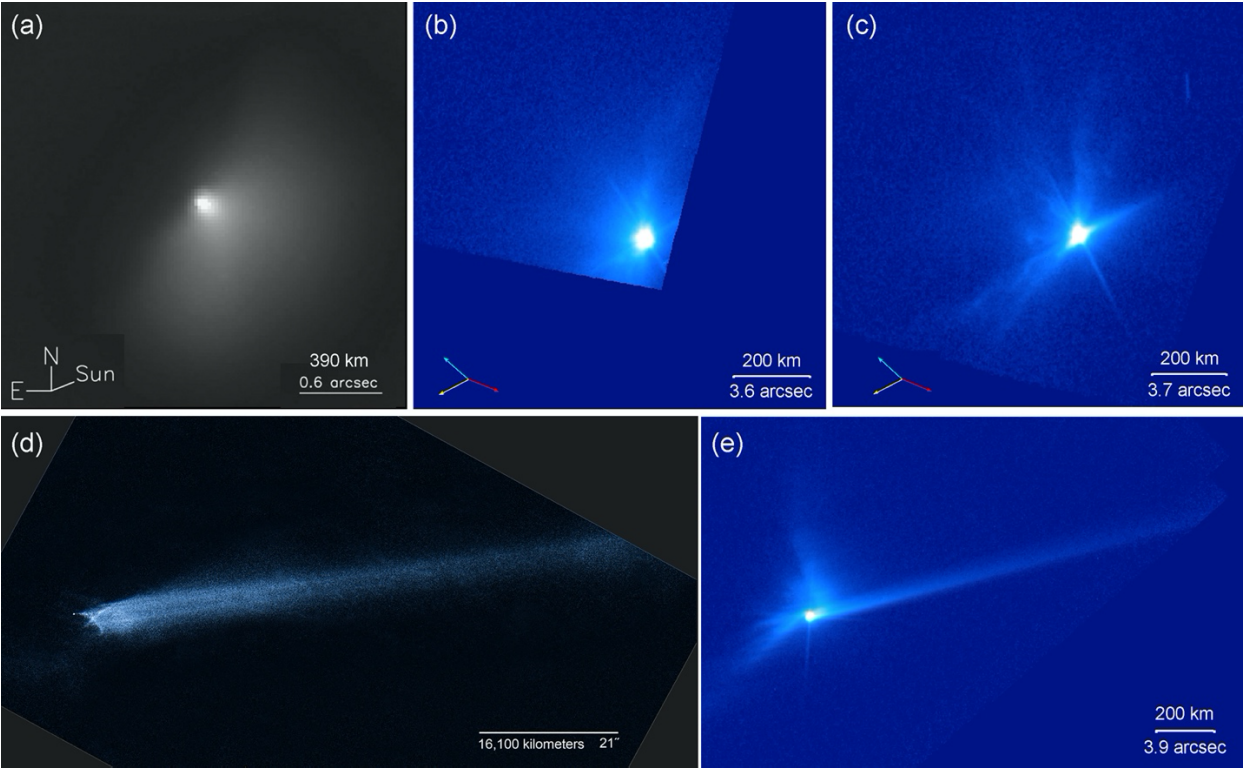

**Extended Data Figure 2.**

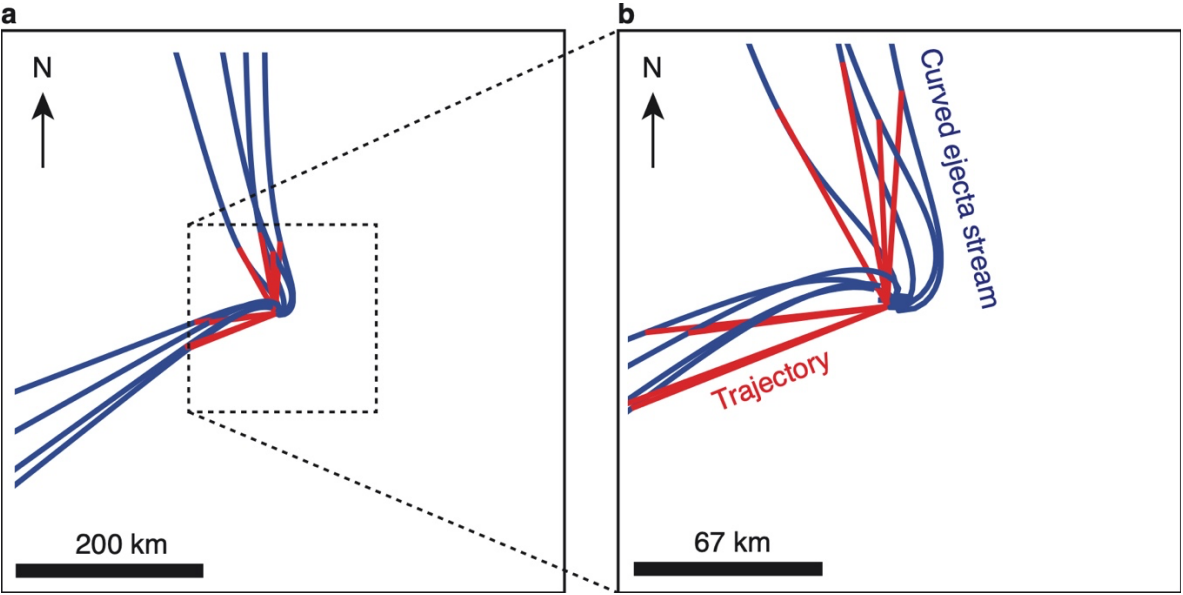

**Extended Data Figure 3.**

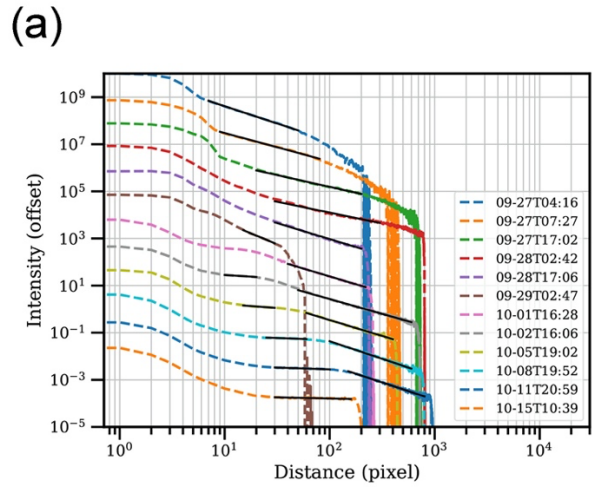 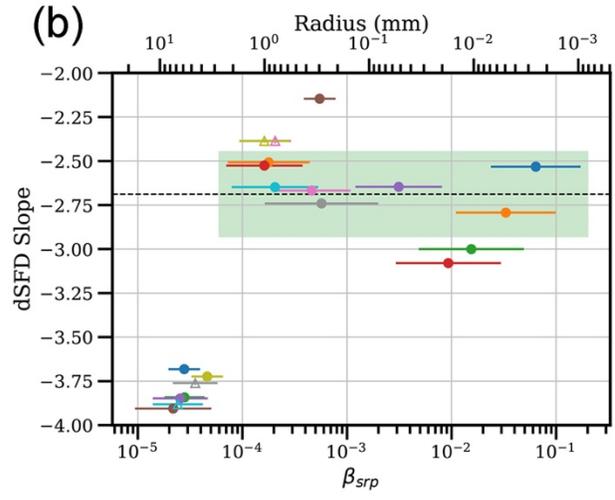

**Extended Data Figure 4.**

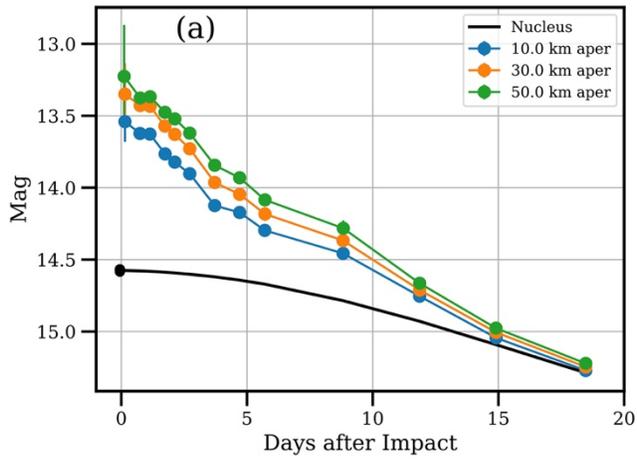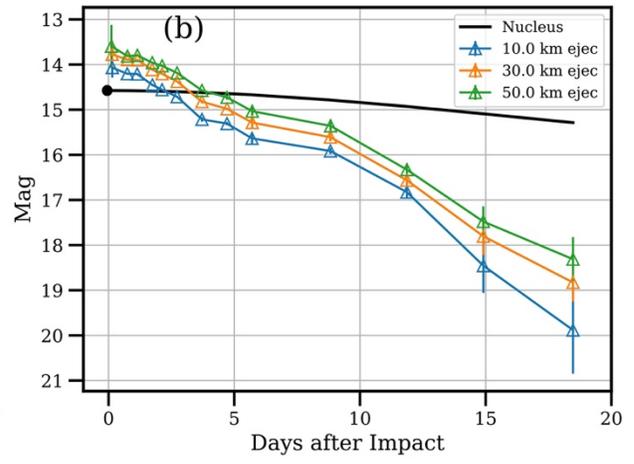

**Extended Data Figure 5.**

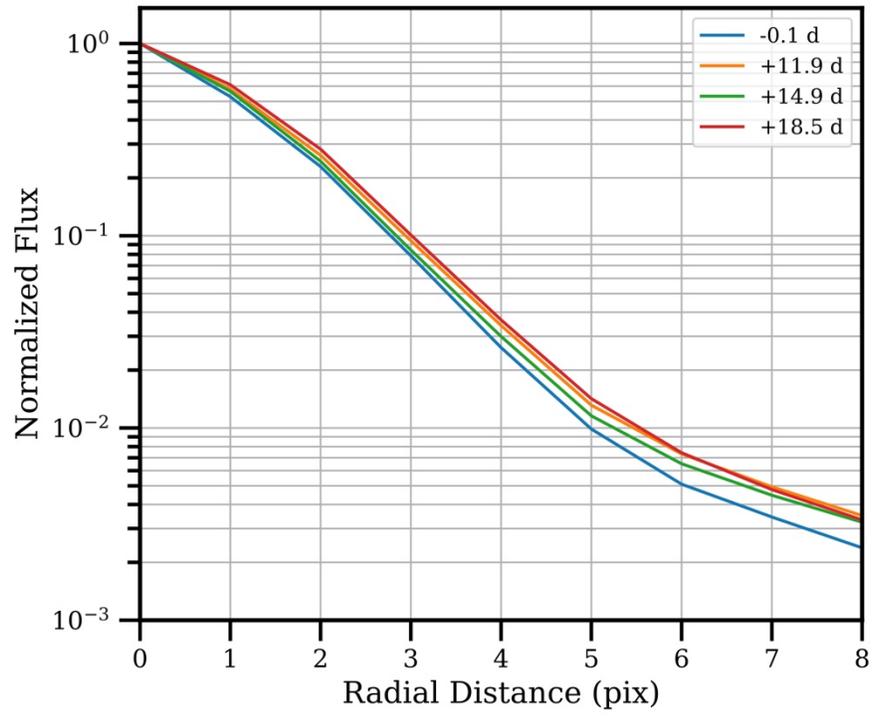

**Extended Data Figure 6.**

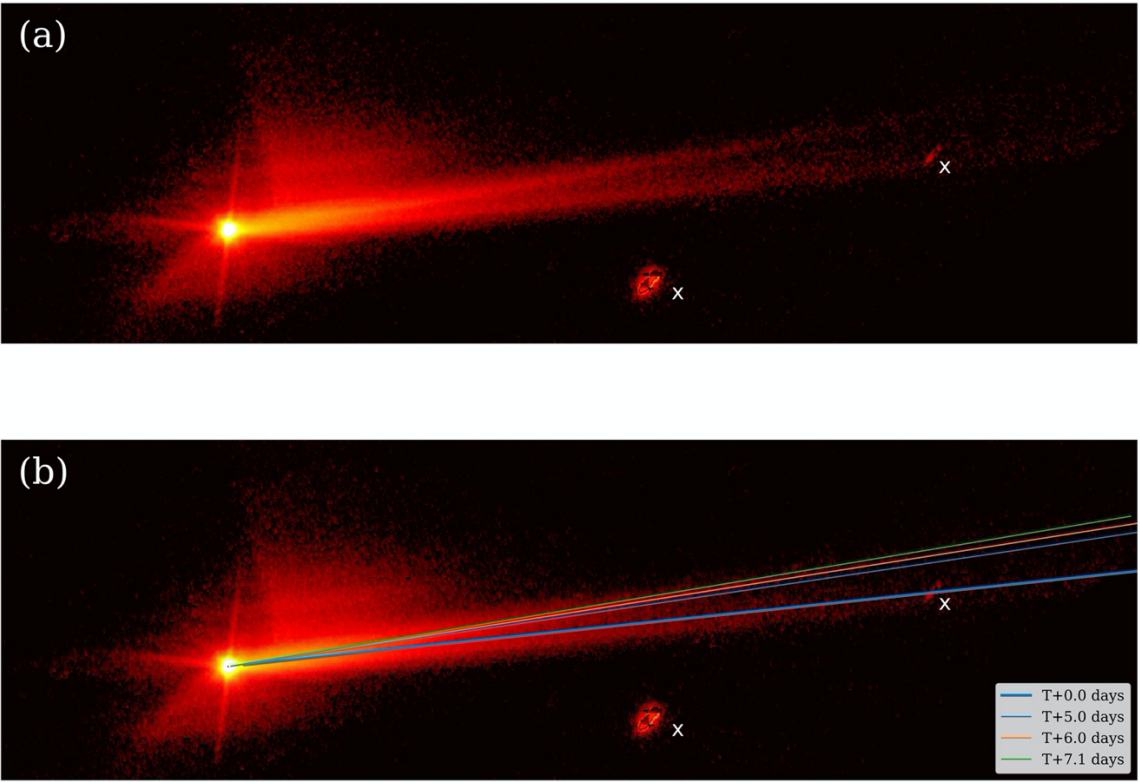

**Extended Data Figure 7.**

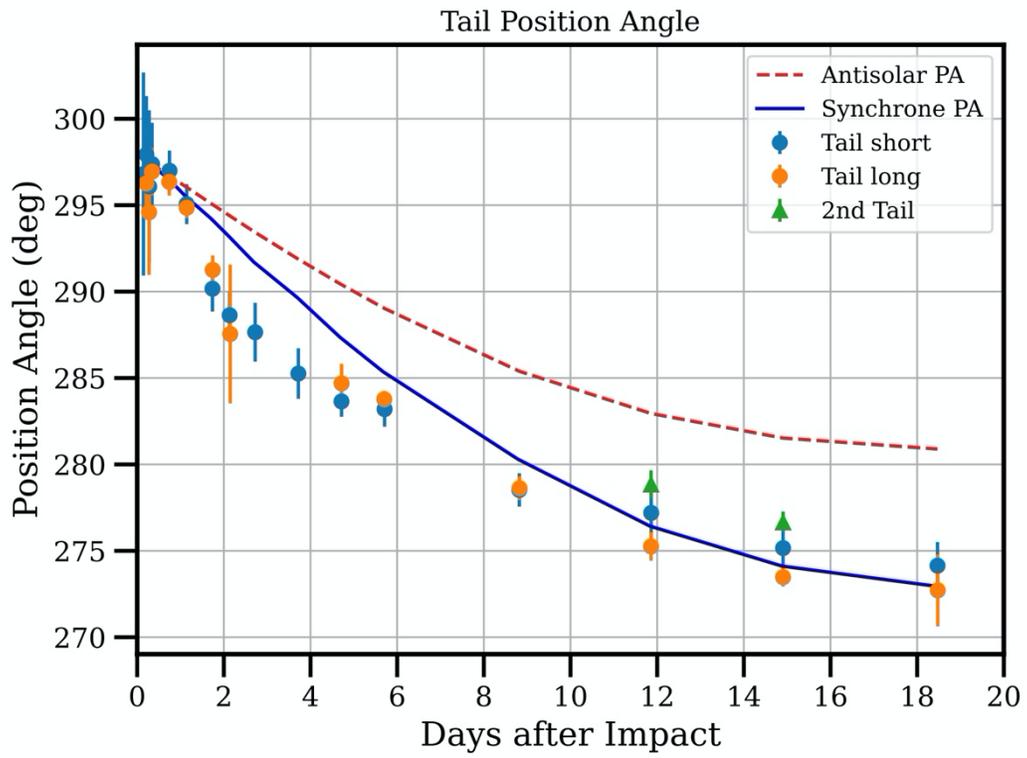